\documentclass[dvipsnames]{aastex61}

\usepackage{myshorthand}
\renewcommand{\revonecolor}{\color{black}}

\newcommand{\sunstardb}{{\tt sunstardb}}

\begin{document}
\title{\sunstardb{}: a database for the study of stellar magnetism \\and the solar-stellar connection}
\author[0000-0002-4996-0753]{Ricky Egeland}
\affiliation{High Altitude Observatory, National Center for Atmospheric Research‡, PO Box 3000, Boulder, CO 80307-3000, USA}
\correspondingauthor{Ricky Egeland}
\email{egeland@ucar.edu}

\begin{abstract}
The ``solar-stellar connection'' began as a relatively small field of
research focused on understanding the processes that generate
magnetic field in stars and which sometimes lead to a cyclic pattern
of long-term variability in activity, as demonstrated by our Sun.
\revone{This area of study has recently become more broadly
  pertinent to questions of exoplanet habitability and exo-space 
  weather, as well as stellar evolution.}  In contrast to other areas
of stellar research, individual stars in the solar-stellar connection
often have a distinct identity and character in the literature, due
primarily to the rarity of the decades-long time series that are
necessary for studying stellar activity cycles.  Furthermore, the
underlying stellar dynamo is not well understood theoretically, and is
thought to be sensitive to several stellar properties,
e.g. luminosity, differential rotation, and depth of the convection
zone, which in turn are often parameterized by other more readily
available properties.  Relevant observations are scattered throughout the
literature and existing stellar databases, and consolidating
information for new studies is a tedious and laborious exercise.  To
accelerate research in this area I developed \sunstardb{}, a
relational database of stellar properties and \revone{magnetic} activity proxy time
series keyed by individual named stars.  The organization of the data
eliminates the need for problematic catalog cross matching operations
inherent when building an analysis dataset from heterogeneous sources.
In this article I describe the principles behind \sunstardb{}, the
data structures and programming interfaces, as well as use cases from
solar-stellar connection research.
\end{abstract}



\section{Motivation}

Observations of the Sun through the telescope began around the time of
Galileo and revealed the presence of sunspots on the solar disk.
Diligent observation of this phenomena in later centuries lead to the
discovery of an $\sim$11 year cycle in the appearance and
disappearance of solar spots \citep{Schwabe:1844}.  \citet{Hale:1908}
identified the magnetic nature of these spots, and therefore of the
solar cycle.  Work has progressed over the subsequent century to
explain the formation of these concentrated regions of magnetic field
in the convective outer envelope of the Sun and their organized
spatio-temporal patterns throughout the solar cycle.  At present, a
magnetohydrodynamic dynamo is presumed to generate the large-scale
magnetic fields responsible for the solar cycle, but the detailed
mechanisms and processes are still in dispute \revone{(see
  \citet{Charbonneau:2010}, \citet{Charbonneau:2014}, and
  \citet{Cameron:2017} for recent reviews)}.  The principal difficulty
in understanding the magnetic Sun is the inability to observe
sub-surface magnetic field, coupled with the computational
intractability of modeling a star from the deep interior all the way
to the observable surface \revone{(see the model domains presented in
  \citet{Charbonneau:2014})}.  This latter difficulty is due to the
steep entropy, pressure, and density gradients near the solar surface
that require high spatial and temporal resolution to model in a direct
numerical simulation.  Finite computational resources therefore force
dynamo models to exclude precisely the layer of the Sun in which
observations are possible.  The observational and theoretical divide
makes the ``solar dynamo problem'' extremely challenging.

\cite{Wilson:1968} recognized that ``if analogous cycles could be
detected in other stars with different values of the fundamental
stellar parameters, the results would be of considerable value in
sharpening the theoretical attack on the whole problem.''  This basic
idea lead to the Mount Wilson Observatory HK Project, a dedicated
observational program of a proxy for stellar magnetism.  The proxy of
choice was emission in the line cores of singly ionized calcium, which
on the Sun is correlated to the presence of regions of enhanced
magnetic field \citep{Skumanich:1975}.  \citet{Wilson:1978} presented results for 91 stars in
the first decade of his synoptic observation program, finding that the
activity of some stars do vary, often in ways very different than the
Sun.  Other stars do not appear to vary much at all.
\citet{Baliunas:1995} analyzed up to 25 years of observations for 111
F, G, and K-type stars, and reported three broad classes of
variability: flat, erratically variable, and cycling.  The cycles are
of varying quality, some of which clearly resemble the solar cycle,
and others that are difficult to determine by eye.

The MWO HK Project also enabled a more direct measurement of stellar
rotation from modulations induced by active regions on the stellar
surface \citep{Baliunas:1983}.  Differential rotation is theoretically
expected to be an important ingredient for the dynamo
\citep[e.g.][]{Parker:1955}.  \citet{Donahue:1996} estimated surface
differential rotation from the seasonal differences in rotation period
measurements, presumably due to active latitude migration as in the
Sun \citep{Donahue:1995}.  Rotation and the convective turnover time,
a function of stellar mass, were shown to be a strongly correlated
with magnetic activity \citep[e.g.][]{Noyes:1984}.

The MWO HK Project initiated a subfield of stellar astrophysics known
as the ``solar-stellar connection'' \citep{Noyes:1996}.  The goals are
to understand the origins of solar and stellar magnetism and their
time-variable patterns.  Research in this area consists of ensemble
studies, comparative studies, and detailed characterizations.
Ensemble studies examine trends in metrics of activity with respect to
one or more independent variables using observations of a collection
of stars.  Comparative studies focus on the detailed similarities and
differences in activity of a small number of well-characterized stars,
one of which is usually the Sun.  Detailed characterizations focus on
a single star, attempting to present measurements of a variety of
fundamental properties to a high degree of accuracy.

Over time, studies in the solar-stellar connection draw on
observations from an increasingly diverse number of sources.
Additional dedicated stellar activity surveys in Ca HK activity were
conducted by the Solar-Stellar Spectrograph at Lowell Observatory
\citep{Hall:1995,Hall:2007b}, the SMARTS HK Program
\citep{Metcalfe:2009}, and the TIGRE project \citep{Schmitt:2014}.
Radial velocity exoplanet searches often produce Ca HK observations as
a byproduct (e.g. HARPS; \citet{Lovis:2011}, California Planet Search;
\citet{Isaacson:2010}).  For these searches, magnetic activity is a
noise source that needs to be understood to avoid false-positives \citep{Queloz:2001,Dumusque:2014}.
Stellar magnetism is also correlated to UV and X-ray emission, making
these observations another useful proxy.  Instruments
producing these observations include \emph{IUE} \citep{Boggess:1978},
\emph{Einstein} \citep{Giacconi:1979}, \emph{ROSAT}
\citep{Voges:1999}, \emph{Hubble Space Telescope}, \emph{XMM-Newton}
\citep{Jansen:2001}, and \emph{Chandra} \citep{Weisskopf:2000}.  See
also the review by \citet{Judge:2012}.
Direct measurements of net surface magnetism and magnetic maps
inferred through Zeeman Doppler Imaging are more rare, but are
becoming available \citep[e.g.][]{Petit:2008,Marsden:2014}.  Finally,
the passage of active regions on a star produces modulations in
visible bandpasses that can be detected high-precision instruments or
methods, such as differential photometry \citep{Lockwood:1997}.
Photometric time series are produced by the Fairborn Observatory
Automated Photometric Telescopes \citep{Henry:1995,Henry:1999}, the
\emph{Kepler} spacecraft \citep{Borucki:2010}, and the upcoming
\emph{TESS} mission \citep{Ricker:2014}.

Results derived from observational time series of activity proxies
include mean values (or other middle value estimates), amplitudes, and
significant periods of variability such as a cycle period or rotation
period.  These derived values are published in the literature, and are
themselves valuable information for further study in conjunction with
estimates of fundamental physical properties such as mass, effective
temperature, radius, luminosity, metallicity, age, etc.  Such
fundamental properties are themselves spread throughout a broad
literature.  A researcher of the solar-stellar connection therefore
spends a considerable amount of time scanning published journal
articles for estimates of the various quantities of interest that may
play a role in determining stellar magnetism and activity.  Modern
bibliographic tools \citep[e.g. ADS][]{Kurtz:2000} and databases like
SIMBAD \citep{Wenger:2000}, VizieR \citep{Ochsenbein:2000}, and
MAST \footnote{\url{https://archive.stsci.edu}} facilitate this a
great deal, but nonetheless the aggregation of relevant information
remains a tedious exercise that is \emph{often repeated by subsequent
  researchers.}  As a Ph.D. student interested in stellar activity I
spent many days manually transcribing tables of results in journal
articles to electronic format that I could use in analysis\revone{,
  and colleagues have certainly done the same.}  Progress in the field can be
accelerated by reducing the duplication of such tedious tasks.

\begin{table}[ht]
  \footnotesize
  \centering
  \begin{tabular}{|ccccc|}
  \hline
  {\bf Magnetic}  & {\bf Structure}    & {\bf Energetics}              & {\bf Dynamo}                  & {\bf Context} \\
  \hline
  var. class      & mass, $M$          & luminosity, $L$               & convection time, $\tau_c$     & age \\
  cycle period    & gravity, $\log g$  & temp., $T_{\rm eff}$          & Rossby num., Ro               & distance \\
  cycle amplitude & radius, $R$        & color, $(B-V)$                & viscosity, $\nu$              & apparent mag. \\
  MWO $S$         & depth of CZ        & spectral irradiance           & mag. diffusivity, $\eta$      & space motion \\
  $R'_{HK}$       & vol. of CZ         & proj. rot.  vel., $v \sin i$  & Reynolds num., Re             & inclination \\
  X-ray, EUV flux & helium fraction, Y & rot. period, $P_{\rm rot}$    & mag. Reynolds num., Rm        & binary? \\
  $|B|$, polarity & metallicity, Fe/H  & diff. rot., $\Delta \Omega$   & Prandtl num., Pm              & planets? \\
  ZDI map         &                    & $E_{\rm magnetic}/E_{\rm kinetic}$   & dynamo num., $C_\alpha$, $C_\Omega$ & \\
  \hline
  \end{tabular}
  \caption{Stellar properties of interest.  CZ is an abbreviation for ``convective zone.''}
  \label{tab:properties}
\end{table}

The \sunstardb{} database was conceived to aggregate observations
relevant to studying the dynamo of the Sun and Sun-like stars into a
single public database.  Table \ref{tab:properties} lists the
quantities of interest that are relevant for \sunstardb{}, categorized
broadly and loosely by type.  The ``magnetic'' properties are the principal data
that define the scope of the database.  \emph{Only stars that have an
observation that is an effect of magnetic field should enter the
database.}  This presently limits the number of objects contained in
\sunstardb{} to order 1000 -- far fewer than contained in the general
purpose database SIMBAD.  The ``structure'' properties pertain to the
structure and composition of the star, many of which are not direct
measurements but are model dependent.  The ``energetics'' properties
are those related to the energy distribution of the star, including
the total luminous output as well as the kinematic motions
(i.e. rotation, convection).  The ``context'' properties are those of
general interest, and are not thought to have a direct consequence on
the physical dynamo mechanism.  Finally, ``dynamo'' properties are
dynamo model-dependent parameters that are somehow related (usually
in a complex and poorly understood way) to the structure and
energetics of the star.

The varied data in Table \ref{tab:properties} exist for several sets
of stars, from many sources, published in various places in the
literature, in on-line databases, or in private custody.  Due to the
complexity of data origin, detailed bookkeeping of data provenance is
necessary in \sunstardb{}.  In many cases there will be conflicting measurements
available, and researchers need to be able to decide which values (if
any) they prefer.  This requires fine-grained data provenance:
i.e. every data point must have an associated source.


\revone{
\sunstardb{} is currently available as ``beta'' release (v0.5.0-beta)
and further development is anticipated.  Source code and examples for
using the database are available on
GitHub\footnote{\url{https://github.com/NCAR/sunstardb}} and a website
describing the current status of the project has been
prepared.\footnote{\url{https://www2.hao.ucar.edu/sunstardb}} I welcome
feedback from the community on the usability of the current
application, as well as ideas for future improvement.
}

This article describes the use cases (Section \ref{sec:usecases}) and
design of \sunstardb{} (Sections \ref{sec:db} and \ref{sec:api}).  The
current status and future work for this project are discussed in
Section \ref{sec:future}.
\section{Use Cases}
\label{sec:usecases}

In order to be a useful tool, \sunstardb{} should simplify several aspects
of solar-stellar connection research.  Below I iterate the use cases
envisioned for the \sunstardb{} service, roughly in order of
importance.  The first four use cases (Sections 2.1--2.4) are at least 
partially implemented, while the last three (Sections 2.5--2.7) are
for future work.  The database schema (Section \ref{sec:db}) has been
designed with each of the following use cases in mind.

\subsection{Add New Measurements to \sunstardb{}}

A researcher has obtained and published new measurements
within the scope of \sunstardb{} (see Table \ref{tab:properties}) and
would like to make these data available to the community.  They fill
out a web form containing and describing their new data, and the
\sunstardb{} curator evaluates the request and updates the database.

\subsection{Discover Stars with Properties of Interest}

A researcher is interested in how (or if) a particular property
depends on another, for example activity cycle period ($P_\cyc$) with
respect to rotation period ($P_\rot$; c.f. \cite{Bohm-Vitense:2007}).
The researcher fashions a query to \sunstardb{} for all stars with a
known $P_\cyc$ and $P_\rot$, and \sunstardb{} returns this data.  The
researcher may additionally specify constraints based on other
properties in order to narrow their study.  For example, they may
provide a spectral type (``only G-type stars'') or color index (``with
$0.59 \leq (B-V) \leq 0.69$'') constraint.

\subsection{Discover Stars with Relevant Time Series}

A researcher is interested in applying a new technique to
estimate cycle periods or cycle quality.  They specify a query to
\sunstardb{} for all stars with $S$-index time series greater than 20
years in length with no gaps greater than 2 years.  \sunstardb{}
returns the set of all time series matching the constraint.

\subsection{Recover Previously Published Relationships}

Suppose a researcher is interested in updating the
\citet{Bohm-Vitense:2007} $P_\cyc$ versus $P_\rot$ diagram with
additional measurements they have made.  They specify query to
\sunstardb{} for stars with $P_\cyc$, and $P_\rot$ that appeared in
\citet{Bohm-Vitense:2007}.  \sunstardb{} returns the data table.

\subsection{Obtain All Known Data for a Particular Star}

A researcher obtained new high resolution spectra for a
particular star and would like to understand these spectra in the
context of the magnetic activity of this star.  They provide the
star's name to \sunstardb{} and receive all known measurements and
time series for that star.

\subsection{Identify a Star in a Diagram}

A researcher uses \sunstardb{} to generate a scatter plot of
activity ($\log(\RpHK)$) versus Rossby number ($Ro$).  They identify
an interesting outlier and want to investigate other characteristics
of the star.  By clicking on the point, the application identifies the
point and provides links to further information.

\subsection{Examine a Star's Position in Multiple Dimensions}

A researcher used \sunstardb{} to make two plots for an ensemble of
stars, $P_\cyc$ vs $P_\rot$ (e.g. \citealt{Bohm-Vitense:2007}) and
$\RpHK$ versus $Ro$.  They have identified an interesting outlier in
the latter plot and would like to know its position in the former.
Hovering the mouse pointer over a point in one plot highlights the
corresponding star in the other plot.

\section{Database Design}
\label{sec:db}

\sunstardb{} is implemented in the PostgreSQL version 9 relational
database.  Relational databases model data as a set of tables (a set
of columns and rows) with a unique ``primary key'' identifying each
row and relationships between tables described through ``foreign
key'' relationships of their constituent columns.  The relational model
eliminates data redundancy; e.g. each star is only stored once in the
database, and all data associated with that star are linked through a
foreign key.  Database queries are phrased in the Structured Query
Language (SQL), and results are efficiently accessed through the use
of database indexes.

\begin{figure}
  \includegraphics[width=\textwidth]{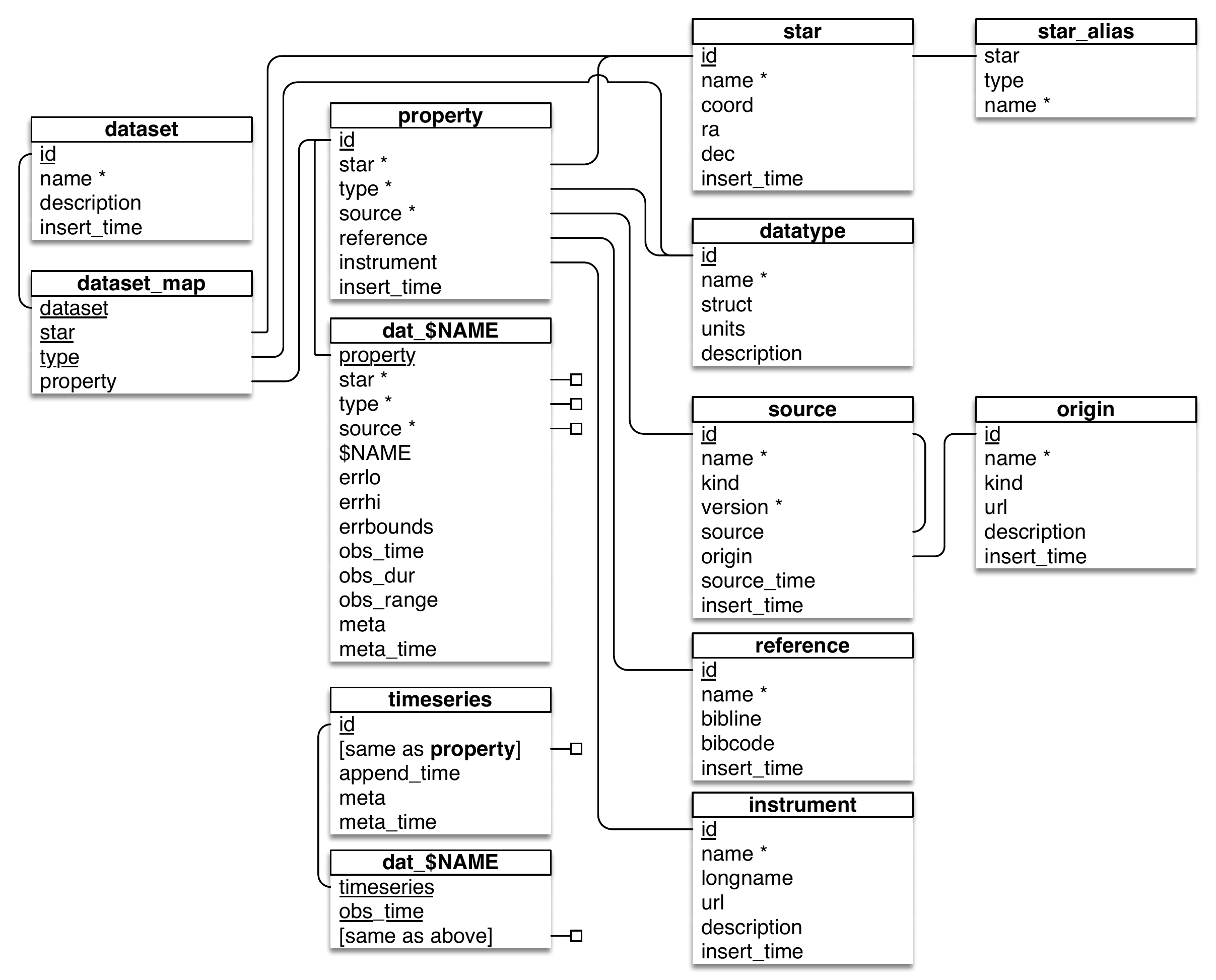}
  \caption{\sunstardb{} database schema.  Each box indicates a table,
    with the columns in that table listed below.  Columns that are
    part of the unique primary key are underlined.  Columns in an
    additional unique key constraint are starred.  Lines indicate
    foreign key relationships.  To simplify the diagram, some foreign
    key lines do not connect to the destination table (with same name
    as source column), but end in a box.  Measurement data are stored
    in the {\tt dat\_\$NAME} tables, where {\tt \$NAME} is the name of
    the measurement type (found in {\tt datatype}) table.  Columns
    are omitted from the {\tt timeseries} and associated data tables
    {\tt dat\_\$NAME} table, which are identical in form to the {\tt
      property} table and its associated data table.}
  \label{fig:schema}
\end{figure}

The relational data model (or schema) for \sunstardb{} is shown in
Figure \ref{fig:schema}.  The principal table is the \texttt{property}
table.  Each \texttt{property} row declares that a measurement of a
given \texttt{datatype} for a given \texttt{star} exists.  Each
\texttt{property} is associated with a \texttt{source},
\texttt{reference}, and (optional) \texttt{instrument}, which declares
the provenance of that datum.  A \texttt{source} is the digital format
of the data immediately before entering \sunstardb{}.  Its
\texttt{kind} is either \texttt{FILE}, for data originating from a
data file (e.g. ASCII text, or FITS), or \texttt{CODE} for data
generated as a result of some analysis process.  Each \texttt{source}
has an an associated \texttt{origin}, which describes where the source
file or code originated.  The \texttt{origin} typically describes a
web site, data repository, or published paper.  The \texttt{source}
table is optionally self-referential, which allows a \texttt{source} to be part
of a versioned hierarchy.  For example, a new calibration of a time
series may be derived from the previous version, or a more raw form.
A property \texttt{reference} is a published journal reference that
explains the datum.  The (optional) \texttt{instrument} describes the
instrument that is ultimately responsible for the measurement.
Properties that are derived from theoretical models (e.g. Rossby
number) or empirical relationships (e.g. age) do not have an
associated \texttt{instrument}.

Measurements of a given data type are stored in template tables,
indicated as \texttt{dat\_\$NAME} in Figure \ref{fig:schema}, where
\texttt{\$NAME} is unique for each table (e.g. \texttt{dat\_S\_mean}
for a table of mean $S$-indices, \texttt{dat\_P\_cyc} for a table of
cycle periods).  While these \texttt{dat} tables have a 1:1
relationship with the \texttt{property} table, the data are stored
separately to allow for different data structures for measurements.
The standard measurement \texttt{dat} table shown in Figure
\ref{fig:schema} contains a single floating-point value
(\texttt{\$NAME} column) with an (optional) error bar (\texttt{errhi}
and \texttt{errlo}).  The \texttt{dat} tables also store an assortment
of optional metadata for each measurement, including the time of
observation (\texttt{obs\_time}), the duration of the observation
(\texttt{obs\_dur}), and the time range of the observation
(\texttt{obs\_range}), which allows for special PostgreSQL queries
that can select data covering a specific point in time.  The meaning
of these time metadata is dependent on the type of measurement.  For
an integrated light observation such as Johnson $V$ it may simply
describe the time and duration of integration.  For an aggregate
quantity such as a mean $S$-index averaged over decades of individual
observations, it may represent the central time and duration of that
parent dataset.

The \texttt{meta} column contains a compound data object encoded in
the JavaScript Object Notation (JSON).  This special feature of
PostgreSQL allows a single column to contain a complex data object.
In \sunstardb{}, this column stores a property-dependent object
consisting of key-value pairs that describe additional information
about the measurement.  For example, a mean $S$-index may have the
number of observations used for the mean stored under the
\texttt{N\_obs} key in the \texttt{meta} column object.  This
ancillary information can be useful in subsequent analysis, but is not
important enough to warrant a first-class \texttt{property}.
PostgreSQL allows the construction of SQL queries that access and
filter results in the compound JSON object.

A \texttt{timeseries} is nearly identical to a \texttt{property},
except that it describes an array of measurements instead of a single
value.  This is reflected by the fact the corresponding \texttt{dat}
tables have the observation time as part of their unique primary key.
Timeseries have array-wide metadata in the \texttt{timeseries.meta}
column, and individual element metadata in the
\texttt{dat\_\$NAME.meta} column.  Note that a timeseries could
equally well be represented as a set of \texttt{property} rows,
however the distinction is appropriate given that timeseries are
generally used in different ways than other scalar properties.

The \texttt{dataset} and \texttt{dataset\_map} tables define a set of
\texttt{properties} for use in a study.  This could be for a single
star, indicating a choice of measurement where duplicate measurements
exist from different sources, or for a stellar ensemble.  For example,
the cycle period and rotation data table of \citet[][(Table 1)]{Bohm-Vitense:2007}
could be compared to the earlier study of \citet[][(Table 1)]{Saar:1999}.  Each 
of these tables can be represented as a \texttt{dataset} and retrieved
from \sunstardb{} to compare or update.

Stars are stored in the \texttt{star} table along with their
position in celestial coordinates.  Stars may have many
different names according to the various catalogs or surveys in which
they were observed.  \sunstardb{} queries the SIMBAD database for a
list of all known identifiers for a star, and stores the list in the
\texttt{star\_alias} table.  The ``default'' name for a star is stored
in the \texttt{star.name} column, and is equivalent to the default
identifier used in SIMBAD.  \sunstardb{} interfaces will be designed to be
name-agnostic: that is, the user should be able to query for data
using any of the star's names, and should be able to specify the
preferred name to display when outputting results.  For example, the
Henry Draper (HD) catalog number is typically used in the
solar-stellar connection literature as most of the well-observed stars
are contained in this catalog of bright stars (e.g. HD 146233).  However, occasionally
special stars constellation-based names (e.g. 18 Scorpii, 18 Sco).
\sunstardb{} needs to be able to recognize any of these names for the
same star.

\section{Application Programming Interface}
\label{sec:api}

The \sunstardb{} application is implemented in Python version 2.7.  At
this time interfaces exist only for direct programmatic remote access
to the database, but in the future web user interfaces are envisioned.
The application programming interfaces (APIs) are broadly divided into
two classes: those responsible for data input, and those responsible
for querying the database.  These interfaces are described in the
following sections.

\subsection{Data Input Interface: Data Modules}

Data relevant to \sunstardb{} exists in a variety of formats.
Examples include ASCII text files in a multitude of formats, XML
provided by existing web services, and FITS format tables.  In
order to insert these data into \sunstardb{}, it must be converted to
a common format and piped to code that can execute the appropriate
SQL statements to insert the data to the database.  Rather than
defining another intermediate file format for this purpose,
\sunstardb{} uses a concept of \emph{data modules}.  Data modules are
Python packages that contain both the input data in its original
format (e.g. text, XML, FITS) as well as a small code snippet to read
that data and organize it into a data structure understood by
\sunstardb{}.  This design pattern reduces redundant code for the
tedious task of parsing and inserting data to the database.

As an example, we will consider the seasonal $S$-index catalog of
\citet{Duncan:1991}, which was transcribed into a digital format and
is available in the VizieR catalog service under the catalog ID
\texttt{III/159A}.  The catalog data file (\texttt{catalog.dat}) was
downloaded from VizieR and placed into a data package under the
subdirectory \texttt{Duncan1991}.  The directory contains two other
files: \texttt{info.json} and \texttt{\_\_init\_\_.py}.  The first
file contains metadata that describes the data provenance, including
bibliographic information on \citet{Duncan:1991}, the source
instrument (MWO-HK), and a description of the data origin (VizieR
catalog \texttt{III/159A}).  The \texttt{\_\_init\_\_.py} file
declares by convention that the \texttt{Duncan1991} directory is a
Python package, and contains $\sim$50 lines of code that reads and
re-formats \texttt{catalog.dat}.  This is done by declaring a
\texttt{DataReader} class, with a \texttt{data} method that is a
Python generator function that returns a sequence of key-value
dictionaries understood by \sunstardb{}.  With these auxiliary files
and code in place, the common \sunstardb{} insertion script need only
be passed the package name (\texttt{Duncan1991}) in order to insert
this catalog into the database.

The data module design pattern reduces the amount of redundant code
through common \texttt{DataReader} base classes specialized to
different classes of input data.  In the above example, the
\texttt{sunstardb.datapkg.TextDataReader} class is employed, which
provides a convenience method for parsing fixed-width text formats
given a specification.  The remainder of the data package merely
reorders parsed lines into a dictionary understood by the \sunstardb{}
data insertion script.

\subsection{Querying and Jupyter Notebook Integration}

Queries to \sunstardb{} are implemented in the \texttt{SunStarDB}
object as methods prefixed with \texttt{fetch}.  Methods exist to
fetch individual rows or all rows of most of the tables in Figure
\ref{fig:schema}, including \texttt{star}, \texttt{datatype},
\texttt{instrument}, \texttt{reference}, \texttt{source}, and
\texttt{origin}.  These methods provide a means of data discovery; that is,
a way for the user to figure out what data exist in \sunstardb{}.
Fetching a set of properties for analysis is accomplished using the
\texttt{fetch\_data\_table()} method.  Here user specifies a
\texttt{dataset} and an array of \texttt{datatype} as input to filter
the result.  All stars with data matching those criteria are returned
to the user.  Time series are accessed with the
\texttt{fetch\_timeseries} method, which accepts a \texttt{datatype}
and a star name, and can be filtered by \texttt{source}.

\begin{figure}
  \frame{\includegraphics[width=\textwidth]{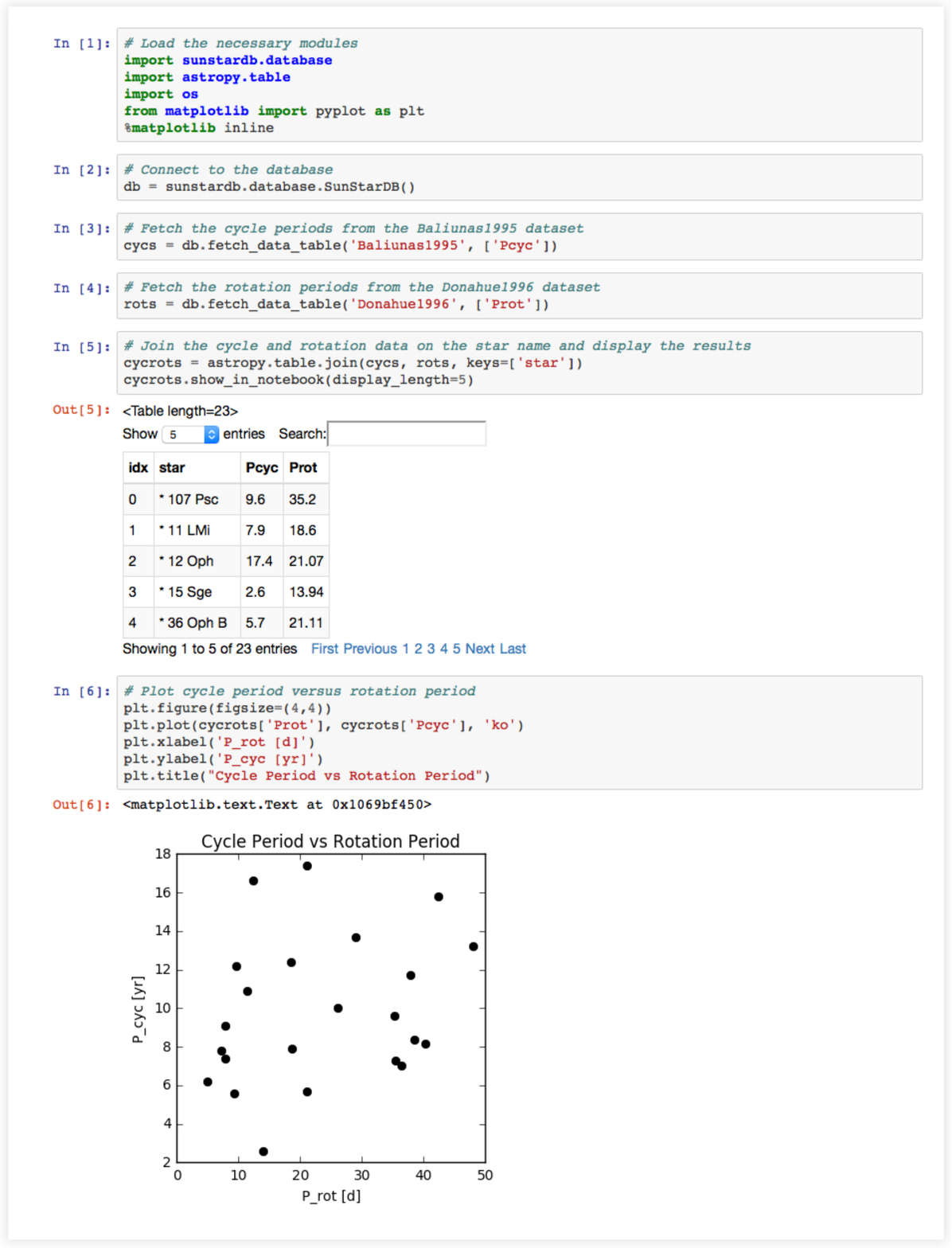}}
  \caption{Example Jupyter notebook session interacting with
    \sunstardb{}.}
  \label{fig:notebook}
\end{figure}

The \texttt{SunStarDB} object has some optimizations for use in the
IPython \citep{Perez:2007} and
Jupyter\footnote{\url{http://jupyter.org}} notebook environment.  An
short example is shown in Figure \ref{fig:notebook}.  Firstly, as with
all python objects in Jupyter, the user has the ability to use tab
completion to discover method names, as well as access to
documentation by appending ``?''  to the method name.  Care has been
taken to provide complete and comprehensible documentation for each
\texttt{SunStarDB} method, including descriptions of method arguments
and output format.  Furthermore, many of the \texttt{fetchall} methods
provided for data discovery return their results as an AstroPy
\citep{Astropy:2013} table object.  By default, this object is
rendered by Jupyter as an HTML table, producing an attractive
presentation of the data.  Calling the \texttt{show\_in\_notebook()}
method of these \texttt{Table} objects adds the ability to paginate,
sort, and filter the table.  For example, typing ``flat'' into the
filter field of a result containing cycle classifications from
\citep{Baliunas:1995} displays only the flat activity stars.  This
functionality can be useful for quickly identifying candidate stars
for specialized studies.

The \texttt{astropy} tables also support SQL-like join operations in
the python environment.  This allows the user to easily ``cross
match'' two result tables by star name, for example (see Figure
\ref{fig:notebook}).  While this functionality is more appropriately
implemented in \sunstardb{} for cases in which the tables are very
large, the catalogs inserted into \sunstardb{} are typically quite
small, of order 100 rows.  In these cases, it can be useful and
intuitive for the user to perform the cross match in the python
environment, where both the input and resulting table can be easily
inspected.

\section{Current Status and Future Work}
\label{sec:future}

\sunstardb{} was developed during my Ph.D. work
\citep{Egeland:2017:thesis} to organize the heterogeneous data I found
useful in the literature.  The current version has rudimentary APIs
for inserting and retrieving data from \sunstardb{}.  Data are
accessed by dataset name, and APIs for retrieving data from multiple
sources are yet to be developed.  Of course, This functionality is
possible for users with direct access to the database and good
knowledge of SQL, but this is not expected to be common for most
users.

Presently, interaction with \sunstardb{} is done from within a Python
programming environment, either through the Jupyter notebook
interface or by custom scripts using the \sunstardb{} package.  A web
user interface is envisioned for simple browsing of the database
contents.  In particular, users should be able to view a web page for
each star that summarizes the measurements and time series available
for that star.  The web application should also provide pages for each
dataset, which can be used to examine the results derived from that
dataset and the provenance of each measurement.

\revone{The current vision for \sunstardb{} is to contain data for every star
which has a measurement that can be used as a proxy for magnetic
activity.}  At the time of this writing \sunstardb{} contains 1191
properties of 6 data types from 12 distinct sources for 233 stars.
\revone{Available data types include mean activity indices $S$ and
$\log(\RpHK)$, as well as rotation periods, variability class
(e.g. ``cycle'', ``flat''), and activity cycle periods.  These are
available for stars ranging from F-type to M-type.  The stars are
mostly on the main-sequence, with a smaller number subgiant and
giant stars.  The sample selection is a reflection of research
priorities in the solar-stellar connection to date.}  It furthermore
contains 1393 time series from 7 distinct sources for 1287 stars, the
vast majority of these from the published table of seasonal mean
$S$-indices by \citet{Duncan:1991}.  The current contents are far from
complete in terms of the vision for \sunstardb{}, but they sufficient
to explore the functionality of the database and even to perform
studies of activity, rotation, and cyclic (or other) variability.
\revone{Standard properties such as spectral type, $B-V$ color,
$T_\eff$, or absolute magnitudes are not yet incorporated into
\sunstardb{}, but are readily available from existing stellar
databases such as SIMBAD or VizieR using the \texttt{astroquery}
package\footnote{\url{https://astroquery.readthedocs.io}}.
Incorporation of these data into \sunstardb{} for all stars with
existing magnetic properties is a near-term priority.}

Future API development will introduce a more widespread
object-oriented approach.  The data provenance information (source,
origin, instrument, etc.) is particularly complex and well suited to
an object representation, rather than the column-like organization
returned by the APIs at present.  \revone{This will make it easier for
the future web application to present provenence information, for
example as a tool-tip when hovering over a data point or table
cell.}  Improved querying is also a high priority, allowing for more
complicated filtering and retrieval of heterogeneous data sets.  These
improvements are focused on the principal purpose of \sunstardb{}: to
simplify data discovery and provide fine-grained meta-data on data
provenance.

\sunstardb{} is currently an unfunded side-project of the author, and
future improvements will likely come in intermittent bursts.  The
vision for \sunstardb{} is to become a standard resource for the
solar-stellar connection community, much like the SIMBAD and ADS
services are for the broader astronomical community.  The current
version described here is only the first step in what will hopefully
be a growing development project.  Feedback on the existing
implementation and ideas for future development will be greatly
appreciated.

\vspace{2em}

\revone{
Thanks to Kim Nesnadny, Travis Kuennen, and Don Kolinski at the High
Altitude Observatory (HAO) for helping to deploy \sunstardb{}.  Thanks
to the anonymous referee for the useful comments which improved this
manuscript.  This research has made use of the SIMBAD database and
VizieR catalog access tool, operated at CDS, Strasbourg, France.
Thanks to Willie Soon and Sallie Baliunas for providing data from the
Mount Wilson Observatory HK Project.  Thanks to Jeffrey C. Hall for
providing data from the Solar Stellar Spectrograph at Lowell
Observatory.  Thanks to Philip G. Judge for organizing the 2014
solar-stellar workshop at HAO which was the genesis of this project.
Thanks to Jeff Hall, Alexei Pevtsov, Travis Metcalfe, Steven Saar,
Christoffer Karoff, Piet Martens, Phil Judge, and Scott McIntosh for
the useful discussions on ideas for a solar-stellar database.
\sunstardb{} development was made possible through funding from the
Newkirk Fellowship at the NCAR High Altitude Observatory.  The author
acknowledges funding from the NCAR Advanced Study Program Postdoctoral
Fellowship.  The National Center for Atmospheric Research is sponsored
by the National Science Foundation.
}

\newpage

\bibliography{bibdesk-master}
\bibliographystyle{aasjournal_links}

\end{document}